\documentstyle[11pt]{article}

 \def\sect{\section} \textwidth
159mm \textheight 220mm \topmargin 0pt \oddsidemargin 5mm

\def\EQ{\begin{equation}} \def\EN{\end{equation}}
\def\bea{\begin{eqnarray}} \def\ena{\end{eqnarray}}

\newcommand{\vs}[1]{\vspace{#1 mm}}

\def\demi{\frac{1}{2}\,}

\newcommand{\ee}{\vspace*{10pt}} \newcommand{\eee}{\vspace*{8pt}}
\newcommand{\deb}{\begin{minipage}{2.5cm} \begin{center} \ee}
\newcommand{\fin}{\eee \end{center} \end{minipage}}
\newcommand{\dd}{\begin{minipage}{0.5cm} \begin{center} \ee}
\newcommand{\ff}{\eee \end{center} \end{minipage}}

 \def\nn{\nonumber \\} \def\pa{\partial}
\def\a{\alpha} \def\b{\beta} \def\c{\gamma} \def\d{\delta}
\def\D{\Delta} \def\e{\epsilon}

  \renewcommand{\t}{\theta}
\newcommand{\shalf}{\frac{1}{2}} 
\begin{document} \begin{titlepage} \begin{center}

\hfill PAR--LPTHE 97--35, OU-HET 273 \\

\vs{25}

{\large\bf A NOTE ON CONSISTENT ANOMALIES}\\

\vs{20}

{\bf Laurent Baulieu}, {\bf C\'eline Laroche}, {\bf Marco
Picco}\footnote{baulieu, laroche, picco@lpthe.jussieu.fr} \\ {\it
LPTHE, Universit\'es Paris VI -- Paris VII, Paris, France}
\footnote{URA 280 CNRS, 4 place Jussieu, F-75252 Paris Cedex 05,
FRANCE.}\\

\vs{5}

{\bf Nobuyoshi Ohta}\footnote{ohta@phys.wani.osaka-u.ac.jp} \\ {\it
Department of Physics, Osaka University, Toyonaka, Osaka 560, Japan}

\end{center}

\vs{20}

\centerline{{\bf{Abstract}}} \vs{5}

Within a BRST formulation, we determine the expressions of the
consistent anomaly for superstrings with extended worldsheet
supersymmetries of rank $N$.  We consider the $O(N)$ superconformal
algebras up to $N=4$, as well as the `small $N=4$' superalgebra. This
is done using a superfield formalism, allowing to recover previous
results that were expressed in components. Moreover, we identify the
`small $N=4$' algebra as the constrained `large $N=4$' via a
self-duality like condition in superspace.

\end{titlepage}

\newpage \renewcommand{\thefootnote}{\fnsymbol{footnote}}
\setcounter{footnote}{0}

\sect{Introduction}

In the last decades, after the work of Ademollo {\it et al.}
\cite{ademollo}, there has been a growing interest in string theories
with extended worldsheet supersymmetries.  More recently, various
theories with different ranks $N$ of local worldsheet supersymmetries
have been shown to be related.  Namely, there are more and more probes
of an embedding of $N$-theories into
$(N+1)$-theories~\cite{BEV,BOP,KST,BGR}.  Therefore, superstring
theories with the higher rank $N$ seem to be privileged.  However, one
crucial point in formulating such highly supersymmetric string
theories is the delicate problem of the superconformal anomaly: its
vanishing is one of the key ingredients to the consistency of a
theory.

There are two different things in the vanishing of the anomaly.  One
thing is the cancelation of the coefficient related to the central
charge of the superalgebra: it gives rise to the well-known critical
dimension of the target space in which the string is embedded for $N
\le 2$.  The second is the so-called consistent anomaly: it is
responsible for the non-conservation of the symmetry generators, or,
put differently, for the breaking of the Ward identities.  Two of the
authors showed in~\cite{BO} that the coefficients of the
superconformal anomalies vanish for all values of $N \ge 3$, but left
quite unexplored the determination of the consistent anomalies.  In
this note, we fill this gap.

The equations that rule the gauge symmetries of superstrings with
local extended worldsheet supersymmetry of arbitrary rank $(N, N')$
were determined in~\cite{BO}.  The (super)-Beltrami parametrisation
\cite{BB} being the most natural parametrisation allowing left/right
factorization, it was generalized to the $N$-extended supersymmetric
case, allowing to present results for the holomorphic sector only (it
all trivially translates to the anti-holomorphic part).

Both classical and ghost conformally invariant multiplets are
assembled into the components of a single $O(N)$-superfield, which
generalizes the Beltrami differential.  This is briefly reviewed in
the next section.  In a path integral formulation, this
`Beltrami-superfield' would be the source for all linear
superconformal generators, and indeed, the BRST symmetry that was
obtained coincides with the one coming from usual operator product
expansion treatments of the superconformal algebra.  Since the BRST
symmetry was found for the complete set of gauge fields and ghosts of
conformal $2D$-supergravity, we are in a position to investigate the
consistent anomalies by means of the descent equations.  It should be
a local functional, with appropriate dimensions and ghost number, and
has to satisfy a Wess and Zumino consistency equation~\cite{BB}.

In this note, we are using an $O(N)$-superfield formalism~\cite{KS}
and we give in section~2 the expressions of the superconformal
consistent anomalies in terms of those superfields, for $N=1,2,3$.
These are in agreement with our former results~\cite{BO}, that were
expressed in components.

Moreover, we have computed the anomalies in the yet unexplored cases
of `small' and `large' $N=4$ supersymmetry~\cite{ademollo, N4}.  We
also indicate how the small $N=4$ superconformal algebra can be
deduced from the large $N=4$, by some self-duality-like constraint
imposed on the $O(N)$-superfield.  This constraint allows one to keep
only half of the fields of the large $N=4$ and is quite analogous to
the reduction in~\cite{ademollo} from $32$ to $16$ superconformal
generators.  This is presented in section~3, and compared with the
results we obtain using operator product expansion techniques.

Finally, we present the intuitive fact that no consistent {\it local}
expression of the anomaly can exist for $N>4$.  In fact, giving a
sense to string theories with higher supersymmetries than $N=4$
appears to be quite delicate: some generators of the superconformal
algebra have negative weight and thus can not be physical.

\sect{Consistent anomaly for the $O(N)$ superconformal algebra}

\subsection{Conventions and notations}

As described in \cite{BO}, one can unify, in the Beltrami
parametrisation, the $O(N)$ gauge superfield $M^z$ and superghost
$C^z$ into a single superfield $\hat{M}^z$, graded with respect to the
ghost number: \bea \label{1} M^z &=& dz + {\mu}^z_{\bar{z}} d\bar{z} +
\sum_{p=1}^{N} \frac{1}{p !}\,
 \t^{i_1}\t^{i_2} \cdots\ \t^{i_p} \,(m^z_{\bar{z}})_{i_1 i_2 \cdots
i_p}
 d\bar{z}, \\ C^z &=& c^z + \sum_{p=1}^{N} \frac{1}{p !}\,
\t^{i_1}\t^{i_2}\cdots
 \ \t^{i_p} \,c^z_{i_1 i_2 \cdots i_p}, \\
 \hat{M}^z &=& M^z + C^z.  \ena

The differential $d$ and the BRST operator $s$ are unified into a
single graded operator $\hat{d}$ such that the BRST equations for the
gauge symmetry are expressed as a vanishing curvature condition in the
$N$ superspace \bea \hat{d} &=& d + s, \nn \hat{d} \hat{M}^z &=&
\hat{M}^z \pa_z \hat{M}^z
 - \frac{1}{4} \sum_{i=1}^N (D_i \hat{M}^z)^2, \ena with the
superderivative $D_i = \pa_{\t_i} + \t_i \pa_z$ satisfying the
anti-commutation relation \bea \{D_i,D_j\} = 2 \d_{ij} \pa_z, \ena
required for the BRST algebra to close.  Since we only deal with the
holomorphic parts, we will not explicitly display the $z$ and
$\bar{z}$ subscripts or indices anymore, keeping in mind the weights
of the various fields and denoting $\pa_z$ simply by $\pa$.

The full BRST algebra can be obtained from the one for the
superghosts~\cite{BOP,BA,FBO}, {\it i.e.}  \bea \label{s} s C = C\,
\pa C - \frac{1}{4} \,\sum_{i=1}^N \,(D_i C)^2, \ena with the
following correspondence \bea s &\rightarrow& d + s, \nn
\label{tofields} C &\rightarrow& M + C.  \ena Therefore, BRST
transformations will be given for the ghosts (superfield or
components), but the extension to the graded superfield (and therefore
to the gauge superfield) is straightforward.

This same correspondence also enables one to simplify the problem of
finding the consistent anomaly $\D^1_2(c_{i_1 \cdots i_p}, m_{i_1
\cdots i_p})$, \footnote{This is a $2$-form with ghost number 1
according to the usual notation $X^{ghost\ number}_{form\ degree}$.}
as follows.  After gauge fixing, the $m_{i_1 \cdots i_p}$,
$p=0,\cdots,N$ ($p=0$ gives the Beltrami $\mu^z_{\bar{z}}$) become
sources for the various currents in the functional integral. The
anomalous Green functions for these currents are then obtained from
suitable differentiation of the broken Ward identities for the
effective action $\Gamma$ (see \cite{BB}): \bea s(\Gamma) = \chi \int
d^2x \ \D^1_2(m,c), \ena where the coefficient $\chi$ has to vanish
for the superstring theory to be consistent. For $N \le 2$, imposing
$\chi = 0$ determines the dimension of the target space, whereas for
any $N \ge 3$, the coefficient $\chi$ has been shown to be
zero~\cite{BO}.

Since $s^2=0$, the consistent anomaly is constrained by the following
Wess-Zumino consistency condition \footnote{ Indeed, $s$ and $d$
respectively increase the ghost number and form degree by one.}  \bea
\int d^2x \ s\D^1_2 = 0 \qquad \rm{ or } \qquad s\D^1_2 = - d \D^2_1 ,
\ena and the problem of finding the consistent anomaly thus reduces to
a question of finding the cohomology of $\hat{d}$ provided one defines
a graded object $\hat{\D} = \D_0^3 + \D_1^2 + \D_2^1$ using the
descent equations \bea s\D^1_2 + d \D^2_1 = 0, \nn s\D^2_1 + d \D^3_0
= 0, \nn s\D^3_0 = 0.  \ena Here, the correspondence (\ref{tofields})
allows a great simplification: instead of the cohomology of $\hat{d}$,
one can rather solve the cohomology of $s$ at ghost number $3$ (up to
total derivatives since $\hat{d}=d+s$ and since a total derivative in
$\D_0^3$ always translates into a total derivative in $\D_2^1$ and
thus does not contribute).  So the object we should look for is the
$3$-ghost $0$-form $\D_0^3(c_{i_1\cdots i_p})$ that is BRST closed, up
to total derivatives and $s$-exact terms (we will also refer to
$\D_0^3$ as to the consistent anomaly). The last step is the
correspondence~(\ref{tofields}) that enables us to deduce the physical
consistent anomaly $\D_2^1$.

\subsection{Superfield formalism}

The BRST algebra or the consistent anomaly can be computed within a
component formalism, where one extracts the transformations for the
various $c_{i_1 \cdots i_p}$ ghosts as the $\t^{i_1} \cdots \t^{i_p}$
terms. On the other hand, on supersymmetry grounds, the whole
calculation can be done using a superfield formalism if one notices
that $c_{i_1 \cdots i_p}$ and its transformation are the lowest
component of the superfields $D_{[i_1}\cdots D_{i_p]}C$ and $s
D_{[i_1}\cdots D_{i_p]}C$ (where the brackets mean antisymmetrisation
of the indices). The latter can be obtained from (\ref{s}) using the
fact that $s$ and $D_i$ anti-commute.  This formalism exhibits the
statistics and weights of the various ghosts, with $C$, $D_i$
anticommuting and having respective weights $-1$, $1/2$.

In the same way, we will obtain the consistent anomaly $\D_0^3$ as the
lowest component of an anomaly superfield in $N$-superspace, which is
the unique completion of the $(N-1)$ anomaly superfield by terms such
that the whole superfield is $s$-closed, up to total derivatives and
$s$-exact terms. The lowest component of this anomaly supermultiplet
is the superconformal anomaly we are looking for, but the
interpretation for its supersymmetric partners is still not clear.

In the case of $N=0$ supersymmetry, the only $0$-form with ghost
number $3$ which is $s$-closed (its transformation under (\ref{s})
actually gives exactly zero and not a total derivative) is \bea
\label{zero} \D_{N=0} = C\, \pa C \, \pa^2 C, \ena where $C$'s lowest
component is related to the reparametrisation ghost $c$.  Using the
aforementioned correspondence (\ref{tofields}), one can deduce the
$1$-ghost $2$-form consistent anomaly in terms of the Beltrami
differential and extract the expected $\pa_z^3 \mu_{\bar{z}}^z$ term,
violating the conservation law of the energy momentum tensor.

Going up to $N=1$ supersymmetry on the worldsheet, one now has a
non-vanishing $DC$ whose lowest component is the holomorphic part of
the supersymmetric ghost, $\c$. Generalizing (\ref{zero}) to the
supersymmetric case requires adding all possible $0$-form, $3$-ghost
terms made out of $C$ and $DC$ and selecting the BRST-closed
combination.  This leads to \bea \label{one} \D_{N=1} = C\, \pa C \,
\pa^2 C - C\,(\pa D C)^2 + \demi\pa C\,DC\,\pa DC, \ena whose
$s$-transformation is again zero. The lowest component of~(\ref{one})
is in agreement with the results in~\cite{BO} and the $\t$-term is
found to be $s(\pa c \pa \c)$ so that it has no physical significance
whatsoever.

The generalization to higher $N$ worldsheet supersymmetries is
identical: one considers all terms containing the non-vanishing
$D_{[i_1}\cdots D_{i_p]}C$ for $p = 0, \cdots , N$ and extracts the
appropriate combination.

Here, the $O(N)$ structure becomes natural, and the $N=4$ bound as to
the existence of a consistent anomaly becomes manifest.  Indeed, let
us consider a generic term occurring in $\D_0^3(C)$.  Since we are
interested in terms that contribute to $\D_2^1(c, m)$, we should
examine terms of the form\footnote{In (\ref{gen}), the $C$ is needed
to extract a $dz$ (we want to extract the $dz d\bar{z}$ term).  The
two other superghosts $D_{[i_1}\cdots D_{i_{p}]}C$ are the same,
because starting from $\D_2^1(c, m)$, one has to differentiate with
respect to $c_{i_1 \cdots i_p}$ and $m_{i_1 \cdots i_p}$ to obtain the
anomalous conservation law for the corresponding generator.}  \bea C\
\pa^{a_1}D_{[i_1}\cdots D_{i_{p}]}C\ \pa^{a_2}D_{[i_1}\cdots
D_{i_{p}]}C, \label{gen} \ena with $p \leq N$. To see what happens
when $N$ increases, we focus on the $p=N$ term. For this particular
term, the total number of $\pa_z$ derivatives $n = a_1 + a_2 =
3-N$. Indeed, $n=3-p$ since the whole term has weight~$0$ and $C$,
$D_i$ have respective weights $-1$ and $\frac{1}{2}$. When we move
from $N$ to $N+1$, $n \rightarrow n-1$ until $n=0$ for $N=3$ so that
one would not expect to increase $N$ any further without encountering
problems of locality.

However, moving to $N=4$, one can notice that the `new' superghost
$D_1D_2D_3D_4 \, C$ of weight~$+1$ can be redefined as the derivative
of a weight~$0$ ghost thanks to the fact that its BRST transformation
is a total derivative.  (This is due to the fact that the $O(4)$
superalgebra has a weight~$0$ generator which is related to the
weight~$1$ generator of the $U(1)$ symmetry.  This generator is to be
associated with gauge field and ghost of respective weights~$-1$,
$0$.)

The $N=4$ anomaly can then be found as a local expression in terms of
this weight~$0$ superghost that we will simply denote as
$\pa^{-1}\,D_1D_2D_3D_4 \, C$.

Defining the notation $\tilde{\D}_N$ for $\D_N$ with indices $i,j,k,l$
running from $1$ to $(N+1)$ instead of $N$, the anomaly superfields
can be expressed as follows: \bea \D_{N=2} &=& C\, \pa C \, \pa^2 C -
C\,(\pa D_i C)^2 + \demi\pa C (D_iC) (\pa D_iC) \nn &+& \demi
C(D_{[i}D_{j]}C)(\pa D_{[i}D_{j]}C) + \demi D_iC (\pa D_j C)
 (D_{[i}D_{j]}C), \\ \D_{N=3} &=& \tilde{\D}_{N=2} \nn &-&
C\,(D_{[i}D_jD_{k]}C)^2 + \demi (D^{[i}D_jC)(D^jD_kC)(D^{k]}D_iC) \nn
&+& \demi(D_{[i}D_jD_{k]}C)(D_{[i}D_{j]}C)(D_kC), \\ \label{big4}
\D_{N=4} &=& \tilde{\D}_{N=3} \nn &+& \left[
\demi(D_{[i}D_jD_{k]}C)(D_lC) - C (D_{[i}D_jD_kD_{l]}C) \right]
 \,(\pa^{-1} D_{[i}D_jD_kD_{l]}C ) .  \ena We have checked by
inspection, using weight and ghost number constraints, that these
expressions of the anomalies are the most general ones (up to
$s$-exact terms that we do not mention since they have no physical
significance). For $N \leq 3$, the $s$-transformation of the anomaly
superfields is $0$, but $\D_{N=4}$ transforms into a total derivative.
The additional part for $N=4$ happens to be the aforementioned
weight~$0$ superghost times its BRST transformation: \bea \D_{N=4}
&=&\tilde{\D}_{N=3} - s(\pa^{-1}D_{[i}D_jD_kD_{l]}C )\, (\pa^{-1}
D_{[i}D_jD_kD_{l]}C ).  \ena All these results as well as the BRST
transformations can be re-expressed in components for the ghosts or
gauge fields. The lowest components of the former anomaly superfields
give the consistent anomalies in terms of ghosts that were found
in~\cite{BO} for $N \leq 3$ (eq.~(39) therein).

In the following, we will present our results in components for the
$N=4$ case. To obtain the physical consistent anomaly $\D^1_2$, one
should first use the correspondence (\ref{tofields}) to write
$\D^3_0(M^z+C^z)$ from $\D^3_0(C^z)$, and then extract the
$dzd\bar{z}$ term with a single ghost.  This is done explicitly for
$N=4$ in the next subsection.

\subsection{Component formalism for the $N=4$ case}

The gauge fields and ghosts will be denoted as \vs{5} \begin{center}
\begin{tabular}{|c||c|c|c|c|c||c|c|c|c|c|} \hline \deb fields \fin &
\dd $\mu$ \ff & \dd $\alpha$ \ff & \dd $\rho$ \ff & \dd $\varphi$ \ff
& \dd $\beta$ \ff & \dd c \ff & \dd $\gamma$ \ff &
 \dd $\varrho$ \ff & \dd $\d$ \ff & \dd b \ff \\ \hline \deb
statistics \fin &$+$ & $-$ & $+$ & $-$ & $+$ & $-$ & $+$ & $-$
 & $+$ & $-$ \\ \hline \deb weight \fin & $0$ & $+\frac{1}{2}$ & $+1$
& $+\frac{3}{2}$ & $+1$
 & $-1$ & $-\frac{1}{2}$ & 0 & $+\frac{1}{2}$ & 0\\ \hline
\end{tabular} \end{center} \vs{5} with the following graded fields:
\bea \hat{\mu}\ &=& dz + \mu \, d\bar{z} + c, \nn \hat{\alpha}_i \,
&=& \alpha_i \, d\bar{z} + \gamma_i, \nn \hat{\rho}_{ij} &=& \rho_{ij}
\, d\bar{z} + \varrho_{ij}, \nn \hat{\varphi}_i \, &=& \varphi_i \,
d\bar{z} + \d_i, \nn \hat{\beta} \ &=& \beta \, d\bar{z} + b, \ena
where we use the duals $\tilde{c}_{i_{p+1}\cdots i_{N}}$ rather than
the fields $c_{i_1\cdots i_p}$ themselves if $p \geq N/2$.
\footnote{We define the duals by $c_{i_1 \cdots i_p} = \e_{i_1 \cdots
i_N} \, \tilde{c}_{i_{p+1} \cdots i_N}$. Our normalization for the
totally antisymmetric tensor is $\e_{1 2 \cdots N}=1$.}

Namely our ghost superfield for the large $N=4$ reads \bea C = c +
\t^i \c_i + \demi \t^i \t^j \e_{ijkl} \varrho_{kl}
 + \frac{1}{3!} \t^i \t^j \t^k \e_{ijkl} \d_l +
 \frac{1}{4!} \t^i \t^j \t^k \t^l \e_{ijkl} \pa b.  \ena Note that the
last term in the superghost $C$ is written as the derivative of the
$U(1)$ ghost~$b$ (weight~0). This is how the problem of non-locality
is avoided, thanks to the fact that the `dimension~1 ghost' can be
taken to be $\pa b$.

The BRST transformations for the ghosts derived from (\ref{s}) are
\bea s\, c \, \ & = & c\, \pa c \; - \frac{1}{4} \, \c_{i}^2, \nn s\,
\c_{i} \ & = & c\, \pa \c_{i} \: - \frac{1}{2} \, \c_{i}
 \, \pa c - \frac{1}{2} \, \epsilon_{ijkl} \, \c_{j} \, \varrho_{kl},
\nn s\, \varrho_{ij} & = & c\, \pa \varrho_{ij} + \frac{1}{4} \,
\e_{ijkl} \,
 \c_{k} \, \pa \c_{l} - \frac{1}{4} \, (\d_{i} \, \c_{j}
 - \d_{j} \, \c_{i}) - \frac{1}{2} \, \e_{ijkl} \, \varrho_{km} \,
 \varrho_{ml}, \nn s\, \d_{i} \ & = & c\, \pa \d_{i} \: + \frac{1}{2}
\, \d_{i}
 \, \pa c - \pa \varrho_{ij} \, \c_{j} - \frac{1}{2} \, \pa
 b \, \c_{i} - \frac{1}{2} \, \epsilon_{ijkl} \, \d_{j} \,
 \varrho_{kl}, \nn s\, b \, \ & = & c\, \pa b \; - \frac{1}{2} \,
\c_{i} \,\d_{i}.  \ena

The consistent anomaly for the ghosts is obtained as the lowest
component of (\ref{big4}) and the correspondence (\ref{tofields}) is
used to obtain its expression for the graded fields: \bea \hat{\D} &=&
\hat{\mu} \, \pa \hat{\mu} \, \pa^2 \hat{\mu} - \hat{\mu}
 \, (\pa \hat{\alpha}_i)^2 + \frac{1}{2} \, \pa \hat{\mu} \,
\hat{\alpha}_i
 \, \pa \hat{\alpha}_i + 2 \, \hat{\mu} \, \hat{\rho}_{ij}\,
 \pa \hat{\rho}_{ij} - \hat{\mu} \,(\hat{\varphi}_{i})^2 \nn &-&\demi
\epsilon_{ijkl} \, \hat{\alpha}_i \, \pa \, \hat{\alpha}_j \,
 \hat{\rho}_{kl} + \frac{1}{3} \, \epsilon_{ijkl} \, \hat{\rho}_{ij}
\,
 \hat{\rho}_{km} \, \hat{\rho}_{ml} - \hat{\rho}_{ij} \,
\hat{\alpha}_i \,
 \hat{\varphi}_j + \demi \hat{\alpha}_i \, \hat{\varphi}_i \,
\hat{\beta}
 + \hat{\mu} \, \hat{\beta} \, \pa \, \hat{\beta} .  \ena One then
extracts the $dzd\bar{z}$ terms with one ghost to obtain the physical
anomaly occurring in the broken Ward identities as \bea \D_2^1 = 2\,(c
\,\pa_z^3 \mu - \c_i \,\pa_z^2 \a_i - 2\,\varrho_{ij}
 \pa_z \rho_{ij} + \d_i \,\varphi_i - b\,\pa \b).  \ena

\sect{`Duality' and the small $N=4$}

\subsection{Duality-like conditions towards the small $N=4$}

It is well known that the $O(4)$ superconformal algebra (of central
charge $0$) has two independent $SU(2)$ subalgebras, of canceling
central charges.  These are the so-called `small $N=4$' superconformal
algebras and are obtained by proper truncation of the `large
$N=4$'. In our formalism, the reduction to the small $N=4$ is done by
imposing (anti)self-duality-like conditions on the superfields. It can
be viewed in two different ways: in component formalism, the single
superfield condition \bea \label{duality} \left(D_{[i}D_{j]} + \demi
\e_{ijkl}\,D_kD_l \right)\,C = 0, \ena (or equivalently with a $-$
sign) is compatible with the BRST algebra and allows one to reduce the
independent component ghosts to $c$, $\c_i$ and $c^+_{ij}=\demi
(\varrho_{ij} + \demi \e_{ijkl} \varrho_{kl})$.  Indeed,
(\ref{duality}) written in components gives the relations \bea
\label{dualc} b &=& \pa c, \\ \label{dualg} \d_i &=& \pa \c_i, \\
\varrho_{ij} &=& \demi \e_{ijkl} \, \varrho_{ij} , \ena The latter are
obtained by examining not only the lowest component condition, but all
of them, that is all the different $\t^{i_1}\cdots \t^{i_p}$
terms. These identifications of all components enables us to rewrite
the `dualised' anomaly as the one for the small $N=4$ derived below
simply by $^+\D = 2 \D_{small} $ where \bea ^+\D_0^3 &=& 2 \, [c \,
\pa c \, \pa ^{2} c - c \, (\pa \c)^2
 - \frac{1}{4} \pa ^2 c \, \c ^2 + c \, \varrho_{ij} \, \pa
\varrho_{ij}
 - \c_{i} \pa \c_{j} \, \varrho_{ij} + \frac{1}{3} \, \varrho_{ij}
 \varrho_{ik} \varrho_{kj}], \\ ^+ \D_2^1 &=& 4 \, [c \, \pa^3 \mu -
\c_i \, \pa^2
 \alpha_i - \varrho_{ij} \, \pa \rho_{ij}].  \ena

On the other hand, if one wants to remain in the $O(N)$-superfield
formalism and `dualise' the anomaly superfield (\ref{big4}), then in
addition to (\ref{duality}), two other duality-like conditions are to
be imposed, reproducing (\ref{dualc}, \ref{dualg}) when reduced to
their lowest components.

\subsection{Small $N=4$ superconformal algebra and consistent anomaly}

We have also derived the consistent anomaly for the small $N=4$
superconformal algebra, using usual OPE techniques. Here we present
our results and provide a check of the duality-like constraint, by
identifying the two small $N=4$ anomalies.

The small $N=4$ superconformal algebra consists of stress-energy
tensor $T$, four supercurrents $G_a^\pm$ ($a=1,2$), and $SU(2)$
currents $J_i$ $(i=1,2,3)$. Their respective weights are $2$,
$\frac{3}{2}$, $1$, and their operator product expansions can be
found, for example, in~\cite{ademollo,OO}.  Introducing the
corresponding ghosts $c,\c_{\pm a}, c_i$ (weights $-1$, $- \demi$,
$0$) and antighosts $b,\b_{\pm a}, b_i$, their operator product
expansions are \bea c(z) b(w) \sim \frac{1}{z-w}, \nn \c_{\pm a}(z)
\b_{\mp b}(w) \sim \frac{\d_{ab}}{z-w}, \nn c_i(z) b_j(w) \sim
\frac{\d_{ij}}{z-w}.  \ena One then computes the BRST charge for this
system and asks for its nilpotency, which requires the central charge
of the ghost system to be $c=+12$ so that the matter part must have
negative central charge $-12$ (this corresponds to the well-know
critical `dimension' $-2$ for the target space).  The BRST charge
allows to find the transformations for the ghosts by usual OPE
calculation as \bea s c &=& c\pa c - \c_{+1}\c_{-1}- \c_{+2}\c_{-2},
\nn s \c_{+1} &=& c\pa \c_{+1} - \shalf \c_{+1}c_3 - \shalf \c_{+1}\pa
c
 - \shalf c_1\c_{+2} + \frac{i}{2} c_2 \c_{+2}, \nn s \c_{+2} &=& c\pa
\c_{+2} - \shalf \c_{+1}c_1 - \shalf \c_{+2}\pa c
 + \shalf c_3\c_{+2} - \frac{i}{2} c_2 \c_{+1}, \nn s \c_{-1} &=& c\pa
\c_{-1} + \shalf \c_{-1}c_3 - \shalf \c_{-1}\pa c
 + \shalf c_1\c_{-2} + \frac{i}{2} c_2 \c_{-2}, \nn s \c_{-2} &=& c\pa
\c_{-2} + \shalf \c_{-1}c_1 - \shalf \c_{-2}\pa c
 - \shalf c_3\c_{-2} - \frac{i}{2} c_2 \c_{-1}, \nn s c_1 &=& c\pa c_1
- i c_2 c_3 + \c_{+1}\pa \c_{-2} + \c_{+2}\pa\c_{-1}
 - \pa\c_{+1} \c_{-2} - \pa\c_{+2} \c_{-1}, \nn s c_2 &=& c\pa c_2 + i
c_1 c_3 + i \c_{+1}\pa \c_{-2} - i \c_{+2}\pa\c_{-1}
 - i \pa\c_{+1} \c_{-2} + i \pa\c_{+2} \c_{-1}, \nn s c_3 &=& c\pa c_3
- i c_1 c_2 + \c_{+1}\pa \c_{-1} - \c_{+2}\pa \c_{-2}
 - \pa\c_{+1} \c_{-1} + \pa\c_{+2} \c_{-2}.  \ena The consistent
anomaly, as the BRST invariant zero-form with ghost number 3, is then
found to be \bea \label{small} \D_{small} &=& c \, \pa c \, \pa ^{2} c
- c \, c_{i} \,
 \pa c_{i} - i \, c_{1} \, c_{2} \, c_{3}\nn && - 4 \, c \, ( \pa
\c_{+1} \, \pa \c_{-1} + \pa
 \c_{+2} \, \pa \c_{-2}) + \pa c \, \pa (\c_{+1} \,
 \c_{-1} \, + \, \c_{+2} \, \c_{-2}) \nn && + c_{1} \, (\gamma_{+1} \,
\pa \c_{-2} - \c_{-1} \,
 \pa \c_{+2} + \c_{+2} \, \pa \c_{-1} - \c_{-2} \,
 \pa \c_{+1})\nn && + i \, c_{2} \, (\c_{+1} \, \pa \c_{-2} + \c_{-1}
\,
 \pa \c_{+2} - \c_{+2} \, \pa \c_{-1} - \c_{-2} \,
 \pa \c_{+1})\nn && + c_{3} \, (\c_{+1} \, \pa \c_{-1} - \c_{-1} \,
\pa \c_{+1} - \c_{+2} \, \pa \c_{-2} + \c_{-2} \, \pa \c_{+2}).  \ena
This anomaly can be re-expressed in a more concise form, using the
vector $\Gamma_{\pm} \equiv (\c_{\pm 1},\c_{\pm 2})$ and neglecting
total derivatives, as \bea \D_{small} &=& c\pa c\pa^2 c -c c_i\pa c_i
 - c\left( \pa^2 \Gamma_+^t \Gamma_- + 6 \pa \Gamma_+^t \pa\Gamma_-
 + \Gamma_+^t \pa^2\Gamma_- \right) \nn && - \pa \Gamma_+^t c_i {\bar
\sigma}^i \Gamma_-
 + \Gamma_+^t c_i {\bar \sigma}^i \pa \Gamma_- - i c_1 c_2 c_3, \ena
where the superscript $t$ means transpose of the vector $\Gamma_+$.

The anomaly expressed as (\ref{small}) is to be identified with the
`dualised' anomaly $(^+\D) / 2$ of subsection~$3.1$, provided the
different ghosts are related by 
\bea \c_{\pm 1} = \demi (\c_1 \pm i
\c_2) \ ,\ \c_{\pm 2} = \demi (\c_3 \pm i \c_4), \nn
c_1 = 2 i
\varrho_{14} \ , \ c_2 = - 2 i \varrho_{13} \ , \ c_3 = 2 i
\varrho_{12} \ .  \ena

\sect{Concluding remarks}

According to the weight counting discussed below eq.~(\ref{gen}), one
does not expect to be able to increase N further in order to obtain
local consistent anomalies. When applying the mechanism depicted above
to compute a consistent anomaly for the $N=5$ case, one encounters
serious problems of locality: there is no way to construct a {\it
local} anomaly because the `overweighted' ghosts (weights~$\geq 1$) do
not transform into total derivatives anymore. Thus one would be forced
to consider terms with two negative powers of the $\pa_z$ derivative,
which can hardly be given a sense.

This can be understood from previous knowledge. Indeed, the
superconformal anomaly comes from the central term of the algebra, and
K.~Schoutens showed in \cite{KS} that it is not possible to add a
central extension to an $O(N)$ superconformal algebra for $N \ge 5$.
Moreover, such an algebra would have generators of negative weights
(to which our overweighted ghosts are associated) so that there seems
to be no way to consistently define string theories with
$O(N)$-extended worldsheet supersymmetry for $N > 4$ unless the
negative weighted generators can be made unphysical.

\vs{5}

\noindent {\bf Acknowledgments:} Some of the calculations in this
paper were done by using the OPE package developed by K. Thielemans,
whose software is gratefully acknowledged.

\newpage \newcommand{\NP}[1]{Nucl.\ Phys.\ {\bf #1}}
\newcommand{\PL}[1]{Phys.\ Lett.\ {\bf #1}}
\newcommand{\CMP}[1]{Comm.\ Math.\ Phys.\ {\bf #1}}
\newcommand{\PR}[1]{Phys.\ Rev.\ {\bf #1}} \newcommand{\PRL}[1]{Phys.\
Rev.\ Lett.\ {\bf #1}} \newcommand{\PTP}[1]{Prog.\ Theor.\ Phys.\ {\bf
#1}} \newcommand{\PTPS}[1]{Prog.\ Theor.\ Phys.\ Suppl.\ {\bf #1}}
\newcommand{\MPL}[1]{Mod.\ Phys.\ Lett.\ {\bf #1}}
\newcommand{\IJMP}[1]{Int.\ Jour.\ Mod.\ Phys.\ {\bf #1}}
\newcommand{\JP}[1]{Jour.\ Phys.\ {\bf #1}}
\newcommand{\JMP}[1]{Jour.\ Math.\ Phys.\ {\bf #1}}

\begin{thebibliography}{99} \bibitem{ademollo} M. Ademollo {\it et
al.}, \PL{62B} (1976) 105;
 \NP{B114} (1976) 297.  \bibitem{BEV} N. Berkovits and C. Vafa,
\MPL{A9} (1994) 653.  \bibitem{BOP} F. Bastianelli, N. Ohta and
J. L. Petersen, \PL{B327} (1994) 35;
 \PRL{73} (1994) 1199;\\
 N. Berkovits and N. Ohta, \PL{B334} (1994) 72;\\
 N. Ohta and T. Shimizu, \PL{B355} (1995) 127.  \bibitem{KST}
H. Kunitomo, M. Sakaguchi and A. Tokura, \PTP{92} (1994) 1019.
\bibitem{BGR} L. Baulieu, M. Green and E. Rabinovici, \PL{B386} (1996)
91.  \bibitem{BO} L. Baulieu and N. Ohta, \PL{B391} (1997) 295.
\bibitem{BB} L. Baulieu and M. Bellon, \PL{B196} (1987) 142;\\
 L. Baulieu, M. Bellon and R. Grimm, \PL{B198} (1987) 343;\\
 L. Baulieu, R. Stora, Lectures given at July 1987 NATO ASI, {\it on}
 Non-Perturbative Quantum Field Theory, published in Proceedings of
 the Cargese Summer Institute, 1987 (QC174.45:N2:1987), Plenum Press.
\bibitem{KS} K. Schoutens, \NP{B295} (1988) 634; \PL{B194} (1987) 75.
\bibitem{N4} A. Sevrin, W. Troost and A. Van Proeyen, \PL{B208} (1988)
447; \\
 E. Ivanov, S. Krivonos and V. Leviant, \PL{B215} (1988) 689.
\bibitem{BA} L. Baulieu, \PL{B288} (1992) 59.  \bibitem{FBO}
F. Bastianelli and N. Ohta, \PR{D50} (1994) 4051.  \bibitem{OO}
A. Schwimmer and N. Seiberg, \PL{184B} (1987) 191; \\ N. Ohta and
S. Osabe, \PR{D39} (1989) 1641.  \end{thebibliography}
 \end{document}